\documentclass[angeo]{copernicus}

\begin{document}

\title{Deriving the radial distances of wide coronal mass ejections from elongation measurements in the heliosphere -- Application to CME-CME interaction}

\author[1]{N. Lugaz}
\author[2]{A. Vourlidas}
\author[1]{I.~I. Roussev}

\affil[1]{Institute for Astronomy, University of Hawaii, 2680 Woodlawn Dr., Honolulu, HI, 96822, USA}
\affil[2]{Code 7663, Naval Research Laboratory, Washington, DC 20375, USA}

\runningtitle{Deriving CME radial distances from elongation angles}

\runningauthor{N. Lugaz et al.}

\correspondence{N. Lugaz (nlugaz@ifa.hawaii.edu)}

\received{May 29, 2009}
\revised{August 19, 2009}
\accepted{September 1, 2009}

%\specialissue{Three eyes on the Sun -- multi-spacecraft studies of the corona and impacts on the heliosphere}

%% These dates will be inserted by the Publication Production Office during the typesetting process.

\firstpage{1}

\maketitle

\abstract{
We present general considerations regarding the derivation of the radial distances of coronal mass ejections (CMEs) from elongation angle measurements such as those provided by SECCHI and SMEI, focusing on measurements in the Heliospheric Imager 2 (HI-2) field of view (i.e. past 0.3~AU). This study is based on a three-dimensional (3-D) magneto-hydrodynamics (MHD) simulation of two CMEs observed by SECCHI on January 24-27, 2007. Having a 3-D simulation with synthetic HI images, we are able to compare the two basic methods used to derive CME positions from elongation angles, the so-called ``Point-P'' and ``Fixed-$\phi$'' approximations.
 We confirm, following similar works, that both methods, while valid in the most inner heliosphere, yield increasingly large errors in HI-2 field of view for fast and wide CMEs. Using a simple model of a CME as an expanding self-similar sphere, we derive an analytical relationship between elongation angles and radial distances for wide CMEs. This relationship is simply the harmonic mean of the  ``Point-P''
 and ``Fixed-$\phi$'' approximations and it is aimed at complementing 3-D fitting of CMEs by cone models or flux rope shapes. It proves better at getting the kinematics of the simulated CME right when we compare the results of our line-of-sights to the MHD simulation. Based on this approximation, we re-analyze the J-maps (time-elongation maps) in January 26-27, 2007 and present the first observational evidence that the merging of CMEs is associated with a momentum exchange from the faster ejection to the slower one due to the propagation  of the shock wave associated with the fast eruption through the slow eruption.   
 }

\introduction[Motivation]
With the launches of the two {\it Solar Terrestrial Relations Observatory} (STEREO) spacecraft and the {\it Coriolis} spacecraft in 2006 and 2003, respectively, coronal mass ejections (CMEs) can be, for the first-time, imaged continuously from the solar surface to 1~AU with coronagraphic and heliospheric imagers. The CME on January 25, 2007 was the fastest eruption imaged by the STEREO/Sun-Earth Connection Coronal and Heliospheric Investigation (SECCHI) suite to date \citep[]{Howard:2008}. Although there was a 20-hr data gap in SECCHI coverage at the time of the ejection, these observations provide one of the best available tests for methods aimed at deriving CME dynamics from SECCHI observations for two main reasons. 

First, in contrast to slow ejections which arrive at Earth with speeds comparable to that of the ambient solar wind, a CME with initial speed greater than 1,300~km~s$^{-1}$ should remain faster than the ambient solar wind in the entire HI-2 field of view (FOV). Because most of the models of CME deceleration invoke a ``drag''
 term proportional to the difference between the ejection and the ambient solar wind speeds \citep[]{Cargill:2004, Tappin:2006}, the acceleration profile cannot be well constrained by the analysis of slow CMEs. 

A second reason is the presence of a preceding ejection from the same active region. This ejection was launched 16.5 hours earlier and had a speed of about 600~km~s$^{-1}$. According to previous analyses \citep[]{Lugaz:2009b, Webb:2009, Harrison:2009}, the two eruptions interacted in the heliosphere somewhere between 20$^\circ$ and 30$^\circ$ elongation from the Sun. It is expected that fast shock waves can propagate inside preceding ejections \citep[]{Schmidt:2004, Lugaz:2005b} and merge with the preceding shock waves. However, the variation of the shock speed inside the preceding magnetic cloud(s) is not known precisely. Numerical simulations \citep[]{Lugaz:2005b} have shown it can vary greatly due to the large variation in density, magnetic field and alfv{\'e}nic
speed inside the magnetic cloud. Therefore, a constant or near constant speed cannot be assumed for the January 25, 2007 CME; in fact, most methods tested by \citet{Webb:2009} to explain the measurements, including cone models and numerical simulations, fared quite poorly past 25-30$^\circ$ (see their Figure 7) for at least one of the two observed fronts, although the cone model proved quite accurate in fitting the faster front. The authors noted that ``conversion techniques from distance to elongation may require more work.'' It is the goal of this article to continue this process in an attempt to analyze HI observations better.

\section{SECCHI observations of the January 24-25 CMEs and numerical simulation}
The two successive CMEs of January 24-25, 2007 were initially reported by \citet{Harrison:2008}. At the time, the two STEREO spacecraft were still in close proximity with Earth (within 0.5$^\circ$) and STEREO-A was rolled by about 22$^\circ$ from solar north. Beyond COR-2 FOV, only STEREO-A/SECCHI observed these eruptions originating from an active region behind the eastern limb. The two eruptions were first imaged by COR-1 at 14:03UT on January 24, 2007 and 06:43UT on January 25, 2007. Based on their appearance in coronagraphic images, we determined in \citet{Lugaz:2009b} that they were associated with active region 10940 which was about 20$^\circ$ behind the eastern limb at the time of the first eruption. Due to positions of the {\it Solar and Heliospheric Observatory} (SOHO) and STEREO spacecraft in January 2007, no triangulation of the source region of the eruptions is possible, as was done for later CMEs by \citet{THoward:2008b} for example. 

Based on the time-height profiles of the CMEs in the SOHO/LASCO FOV, and using the same position angle (PA 90) for both CMEs, the speed in the corona of the first CME was 600~km~s$^{-1}$ and it was 1350~km~s$^{-1}$ for the second one. The data gap in SECCHI coverage started after 04:53UT and 09:53UT on January 25, 2007 for STEREO-A and B, respectively and lasted until the start of January 26, 2007. Assuming no deceleration, the two ejections should have interacted during this time. After the SECCHI data gap, two or three bright fronts associated with the eruptions were tracked in HI-2 \citep[]{Harrison:2008, Lugaz:2008b, Lugaz:2009b, Webb:2009}, the first front up to elongation angles of about 55$^\circ$ with HI-2 and up to much larger elongation angles ($\sim 90^\circ$) with SMEI \citep[]{Webb:2009}. SMEI observations could not help during the SECCHI downtime, because the CMEs were inside the SMEI exclusion zone circle of 20$^\circ$ around the Sun. 

We performed a numerical simulation of these ejections with the Space Weather Modeling Framework (SWMF) \citep[]{Toth:2005} using the solar wind model of \citet{Cohen:2007}. The simulation set-up and detailed results have been published in \citet{Lugaz:2009b}. Simulations with the ENLIL model of \citet{Odstrcil:2005} and the HAFv.2 model of \citet{Hakamada:1982} and \citet{Fry:2001} have also been performed and published in \citet{Webb:2009}. Based on the numerical analyses, the fronts observed by HI-2 and SMEI have been associated with the two CMEs, validating the numerical models on one hand and helping the analysis of the complex observations on the other hand. The goal of the current study is to test the existing methods to derive CME radial distances from elongation angles with the help of a 3-D simulation.

\section{Determining CME positions from elongation angles: Testing the existing methods}

\begin{figure*}[t]
\vspace*{2mm}
\begin{center}
\includegraphics[width=16cm]{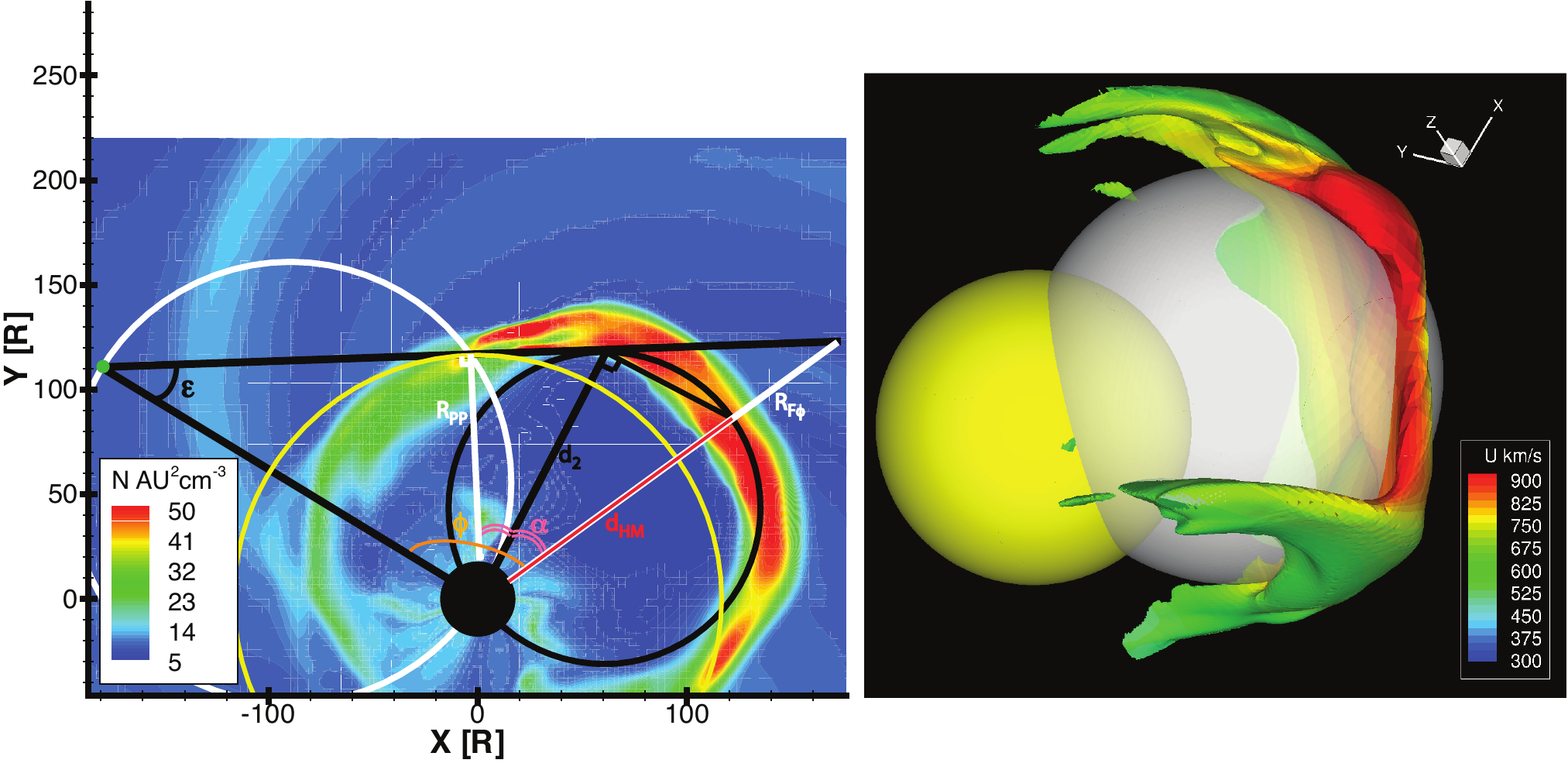}
\end{center}
\label{fig:geom}
\caption{{\it Left}: Geometry of the observations and the methods described in the article. The figure corresponds to the January 24-25 CMEs in the late phase of their merging. This illustrates the different CME positions obtained from one measurement of the elongation angle at $\epsilon$. The black and yellow circles illustrate the model of CMEs used to derive the relation described in the article and the Point-P approximation, respectively; the white circle is the Thomson sphere;  the green dot and black disk are STEREO-A and the Sun, respectively (not to scale). The angle $\phi$ is set at 90$^\circ$ to determine the CME distances but it is shown here as determined from the position of the active region at the start of the eruptions. {\it Right}: The model (expanding propagating sphere) proposed to derive CME position from SECCHI measurements is illustrated for a different simulation (August 24, 2002) with different models of the solar wind and CME initiation. The yellow sphere is centered at the Sun, the white translucent sphere is the model of the CME front and the actual simulated CME is shown as an isosurface of scaled density 20~cm$^{-3}$~AU$^{-2}$ color-coded with the speed.}
\end{figure*}

So far, CME positions have been determined from STEREO observations via 3-D forward fitting of a cone-model or a flux-rope-shaped density enhancement \citep[]{Boursier:2009, Thernisien:2009}, via 3-D reconstruction in COR-2 FOV (i.e. within 20~R$_{\odot}$) \citep[]{Mierla:2008,deKoning:2009}, by mass conservation principles \citep[]{Colaninno:2009}, or by applying one of two simple approximations giving an analytical relation between elongation angles and CME positions \citep[]{Wood:2009, Rouillard:2009, Davis:2009}. These analytical relations provide a quick and easy way to estimate CME dynamics in the heliosphere. 3-D reconstruction and forward modeling are expected to be more accurate than these simple relations, especially in the COR FOV where they have been mostly used so far, but they also have some limitations. For example, the 3-D reconstruction methods require multiple viewpoints, which might become less and less frequent as the STEREO spacecraft separate; when there are multiple observations, they assume that both SECCHI instruments observe the same structure, which is not true in the HI FOV. Additionally, forward modeling attempts to fit geometrical and kinematic information at the same time. To simplify the fit, a kinematic model (often constant acceleration or constant speed) is usually assumed. As noted above, these assumptions cannot be used for complex events, such as those involving CME-CME interactions.

\subsection{The ``Point-P'' and ``Fixed-$\phi$'' approximations}
The intensity of the Thomson scattering depends on the angle between the scattering electron, the Sun and the observer \citep[]{Minnaert:1930}. The loci of the ensemble of points where the intensity of Thomson scattered light is maximum is referred to as the "Thomson surface" \citep[]{Vourlidas:2006}.
In 3-D space, this surface lies on the surface of a sphere with the Sun-observer line as the diameter, and so we refer to this as the Thomson sphere from now on. 
A simple plane-of-the-sky approximation cannot be used with accuracy in the HI FOV \citep[e.g., see][]{Vourlidas:2006}. Therefore, to know which part of a CME is imaged, one needs to consider the complex interaction of the CME 3-D density structure with the Thomson sphere \citep[e.g., see][]{Lugaz:2008b}. An additional problem is that the speed and acceleration should be calculated for the same plasma element (i.e. usually for a single radial trajectory). Even if the CME positions can be determined accurately from HI observations, further assumptions regarding the CME geometry must be made to derive kinematic information, since what is observed over time is not necessarily the same part of the CME, as shown in \citet[]{Lugaz:2009b} and \citet{Webb:2009}.  There are two main simple approximations which have been used to replace the plane-of-sky approximation for heliospheric measurements: they are referred as  ``Point-P'' and ``Fixed-$\phi$'' \citep[]{Kahler:2007,THoward:2007,Wood:2009}; the geometry of the observations and the reconstruction is illustrated in the left panel of Figure~1 for a plot of the simulations of the January 24-25, 2007 CMEs.

The ``Point-P'' (PP) approximation is the simplest possible way to relate elongation angles to CME radial distances while taking into account the Thomson sphere geometry. Assuming a spherical front centered at the Sun, the elongation angle $\epsilon$ and the position of the CME $R_{\mathrm{PP}}$ are related by:  $$R_{\mathrm{PP}} = 
d_{\mathrm{STEREO}} \sin\epsilon,$$ where $d_{\mathrm{STEREO}}$ ($\sim$ 0.97~AU) is the heliocentric distance of STEREO-A for this event. The CME front obtained from this approximation is shown with the yellow circle in the left panel of Figure~1.
Obviously, this approximation is poor for narrow CMEs such as the one studied by \citet{Wood:2009} and for dense streams and corotating interacting regions (CIRs) which are structures of narrow azimuthal extent at 1 AU (the typical width is less than 20$^\circ$ as inferred from \citet{Jian:2006} for example). Even for wide CMEs, the CME fronts are not spherically symmetric, in part due to their interaction with the structured coronal magnetic field and solar wind. This has been shown by multiple-spacecraft observations \citep[e.g.][]{Moestl:2009b} and from simulations \citep[]{Riley:2003, Manchester:2004b, Odstrcil:2005}. Last but not least, the reconstructed CME position is independent of the propagation angle! \citet{Webb:2009} remarked that the PP approximation is not adequate far from the Sun, e.g., in HI-2 and SMEI/camera 2 FOVs.

The ``Fixed-$\phi$'' (F$\phi$) approximation, in turn, takes the opposite philosophy and considers that a single particle, propagating on a fixed-radial trajectory, is responsible for the Thomson scattered light. The elongation angle measurement must simply be ``de-projected'' from the Thomson sphere onto this radial trajectory, resulting in the relation $$R_{\mathrm{F\phi}} = 
d_{\mathrm{STEREO}} \frac{\sin\epsilon}{\sin(\epsilon+ \phi)},$$ where $\phi$ is the angle between the Sun-observer line and the trajectory of the particle. The position obtained from this approximation is noted as $R_{\mathrm{F\phi}}$ in the left panel of Figure~1.
Obviously, this approximation is well adapted for CIRs \citep[]{Rouillard:2008} and small ``blobs'' \citep[]{ Sheeley:2008b,Sheeley:2009,Rouillard:2009}. However, since it assumes that what is tracked is a single point, the method is expected to work poorly for wide CMEs. The equation can be fitted for $\phi$ (assuming no or constant acceleration), giving the origin of the transient \citep[]{Rouillard:2009} and/or the speed. The main limitation of this method is that it completely ignores the CME geometry. It also does not take into account the angle dependency of the Thomson scattering. 

\subsection{Comparison with 3-D simulated data}
We test the two methods with our synthetic line-of-sight procedure and compare the resulting positions to the 3-D simulation for the second CME (January 25 CME) at PA 90.
This work is the continuation of section 4.3 from \citet{Lugaz:2009b} and its associated Figure 6. We derive the elongation angles and radial distances of the CME front as follows: for the line-of-sight images, we use elongation angles measured at the point of maximum brightness at PA 90. 
For the numerical simulation, we use the position of maximum density along different radial trajectories (at longitudes 90$^\circ$, 80$^\circ$ and 70$^\circ$ east of the Sun-Earth line) and on the Thomson sphere, all of these in the ecliptic plane (PA 90). 
Results are shown in the top panel of Figure~2.

\begin{figure}[t]
\vspace*{2mm}
\includegraphics[width=8cm]{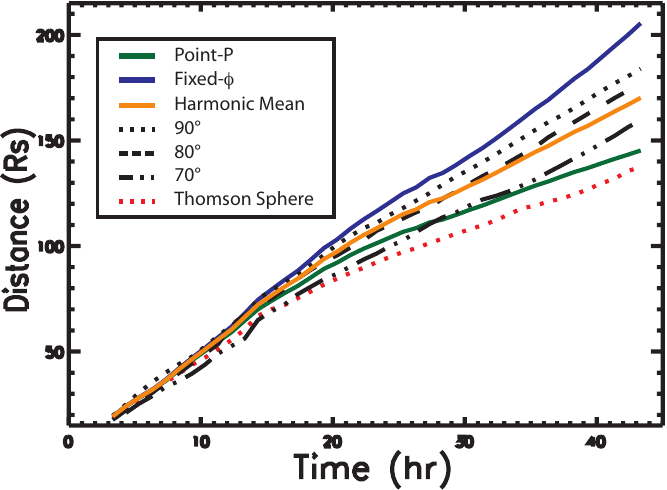}
\includegraphics[width=8cm]{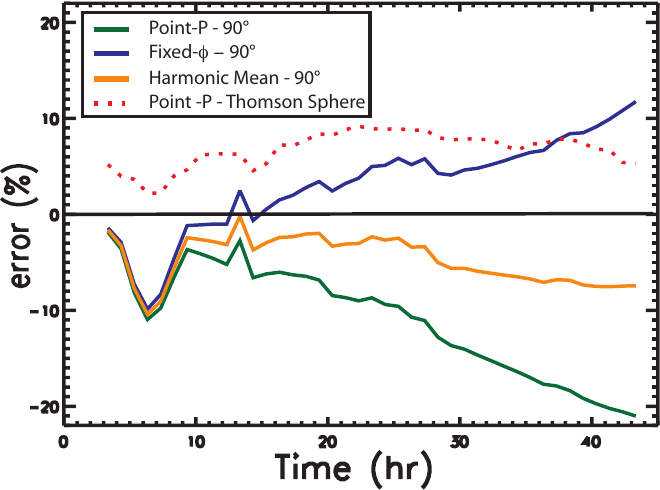}
\includegraphics[width=8cm]{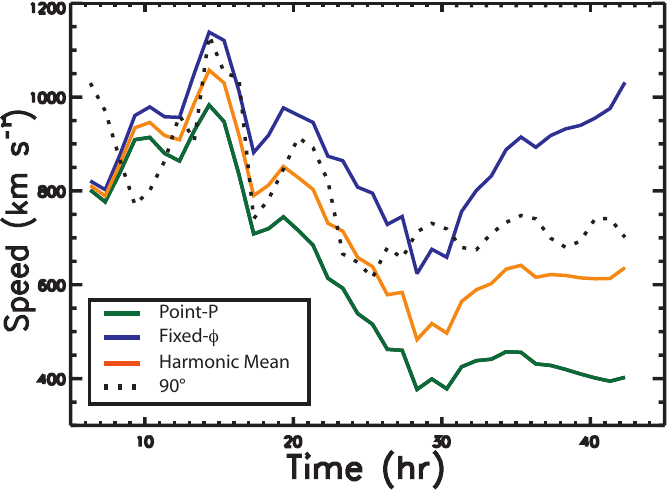}
\caption{Position ({\it top}), error ({\it center}) and speed ({\it bottom}) of the second CME front at PA 90 from the simulation and as derived from the synthetic SECCHI images with the different methods. }
\label{fig:comp}
\end{figure}

Below approximately 100~R$_{\odot}$, the two methods give similar results differing by less than 10$\%$. 
The F$\phi$ method gives slightly better results when compared to the nose of the CME;  the PP approximation works best if one assumes it tracks the intersection of the CME front with the Thomson sphere (see middle panel of Figure~2). Above 100~R$_{\odot}$, the two methods give increasingly different results. Compared to the simulation results along all three radial trajectories presented here, the PP approximation results in a too large deceleration of the CME, whereas the F$\phi$ results in an apparent acceleration. This acceleration appears unphysical, since  CMEs faster than the ambient solar wind are expected to monotonously  decelerate due to a ``drag'' force. 
Similar results have been reported, most recently by \citet{Wood:2009}. The middle panel of Figure~2 shows the errors between the position of the CME front at the limb and the position from each of the two methods, as well as the error between the PP position and the intersection of the CME front and the Thomson sphere. Although the errors are fairly low, they can result in large errors in the velocity and acceleration of the CME (see bottom panel of Figure~2). These methods can provide an average speed of the CME front within the first 100~R$_{\odot}$, but they cannot be relied upon to study complex physical mechanisms such as CME-CME interaction.

\section{Improved method to determine CME position}

Based on the relatively poor results for the PP and F$\phi$ methods, we propose another analytical method based on simple geometric considerations and a simple model of CMEs. We construct this model on a few principles: first, it should take into account the geometry associated with the Thomson scattering as well as the CME propagation and second, it should have the lowest number of free parameters possible. To construct such a model, we start from the knowledge that CMEs are known to evolve self-similarly in the heliosphere \citep[e.g., see][]{Krall:2006}. 
The simplest approximation is to assume that the CME peak density maps out as a sphere connected to the center of the Sun; the center of the sphere propagates in a fixed, radial trajectory (see right panel of Figure~1). In contrast to the PP approximation, the sphere is not centered on the Sun. Consequently, this method takes into account the direction of propagation of the CME. This approximation is also the one used in \citet{Webb:2009} to produce their Figure 1b. 

If we assume no deflection of the CME in the corona or in the heliosphere, the angles defining the trajectory of the center of the sphere can be derived from the flare information (with an understanding of the limitations in the connection between flares and CMEs) or from forward modeling of the COR observations or mass analysis. For the January 25, 2007 CMEs, we will consider that the center propagates from the eastern limb at PA 90. 
There are many ways this sphere ``interacts'' with the Thomson sphere to produce the Thomson-scattered signal. We consider two hypotheses: the geometry associated with the Thomson scattering is dominant and the emission originates from the intersection of the sphere with the Thomson sphere or it is negligible and the emission originates from the line-of-sight tangent to the sphere (see left panel of Figure~1 for the geometry and the notation used). The first hypothesis gives $d_1 = R_{F\phi}$ for the diameter of the circle representing the CME front at the PA where the measurement is made. This PA can differ from the latitude $\lambda$ along which the center of the CME propagates. Correcting for this, the nose of the CME is at a distance of $R_{\mathrm{F\phi}}/\cos(\mathrm{PA}-\lambda)$. This gives a new interpretation for the ``Fixed-$\phi$'' approximation, namely that it gives the diameter of the circle representing the CME at each PA, assuming the emission originates from the intersection of this circle with the Thomson sphere.  

The distance of the point tangent to the CME along the given PA (see the left panel of Figure~1 for the notation) is: $$d_2 = \frac{d\sin\epsilon}{\cos\alpha}= \frac{d\sin\epsilon}{\cos\left(\frac{1}{2}(\phi + \epsilon - \frac{\pi}{2})\right)}\,.$$ The diameter of the circle representing the CME at this PA is simply given by: $$ d_{\mathrm{HM}} = \frac{d_2}{\cos \alpha} = 2\frac{d\sin\epsilon}{1+\sin(\epsilon+\phi)}\,,$$which is the harmonic mean of the PP and F$\phi$ approximations. To obtain the diameter of the sphere, this must also be corrected for the difference between the measured PA and the direction of propagation of the CME:
$$
\frac{1}{R_{\mathrm{HM}}} = \frac{\cos(\mathrm{PA}-\lambda)}{2}\left(\frac{1}{R_{\mathrm{F\phi}}} + \frac{1}{R_{\mathrm{PP}}}\right)\,.
$$
This correction is required because all parts of a CME cannot be assumed to move radially outward with the same speed. Thus, this hypothesis is most likely true for the nose of the CME, which is where the speed must be calculated. 
We plot the position, error and speed derived from this approximation (referred as the harmonic mean (HM) approximation) in the three panels of Figure~2. As can be seen, this simple model gives better results than the PP and F$\phi$ approximations, especially for the speed of the CME at large elongation angles. 

\section{Revisiting the January 26-27, 2007 observations: CME-CME merging}

\begin{figure}[t]
\vspace*{2mm}
\begin{center}
\includegraphics[width=8cm]{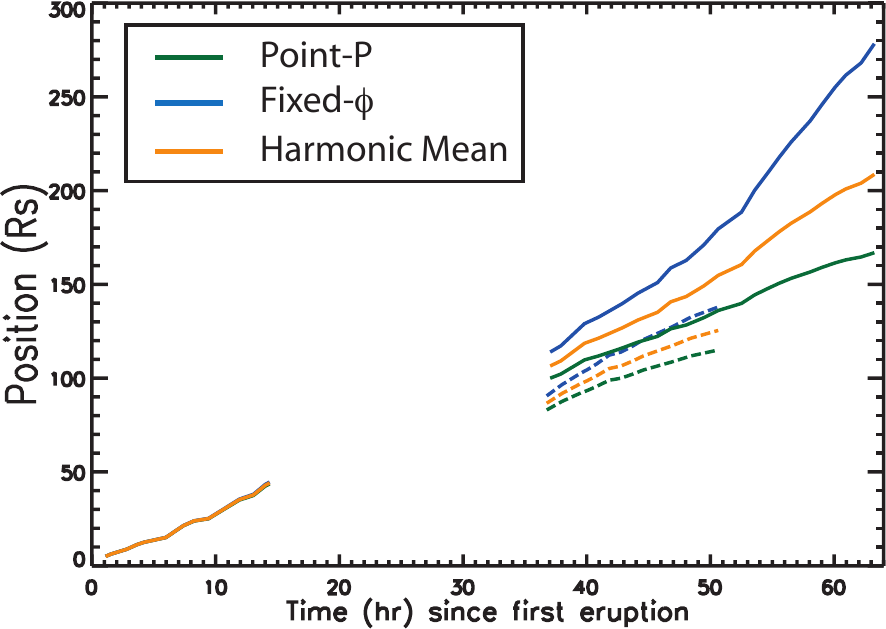}
\includegraphics[width=8cm]{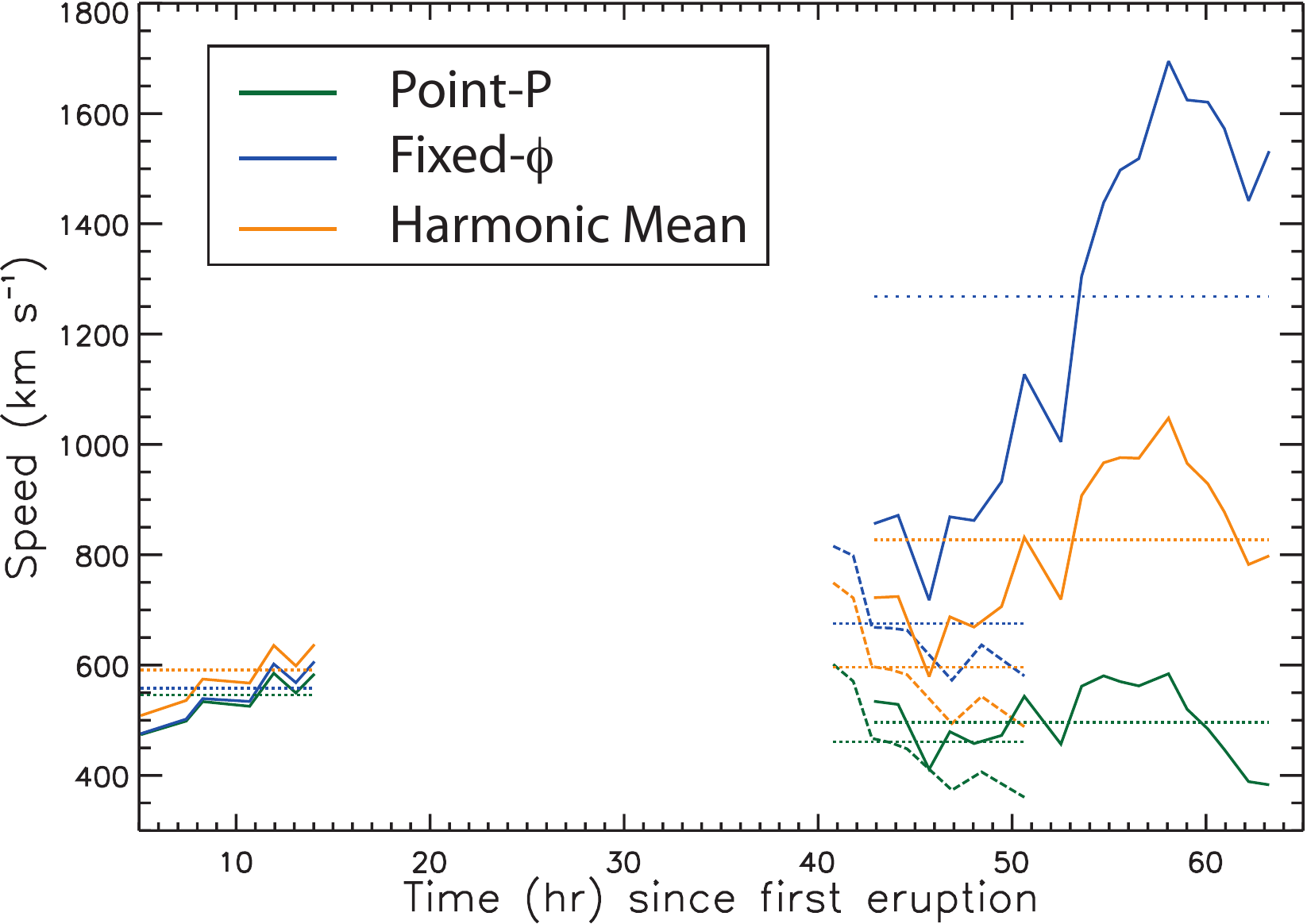}
\includegraphics[width=8cm]{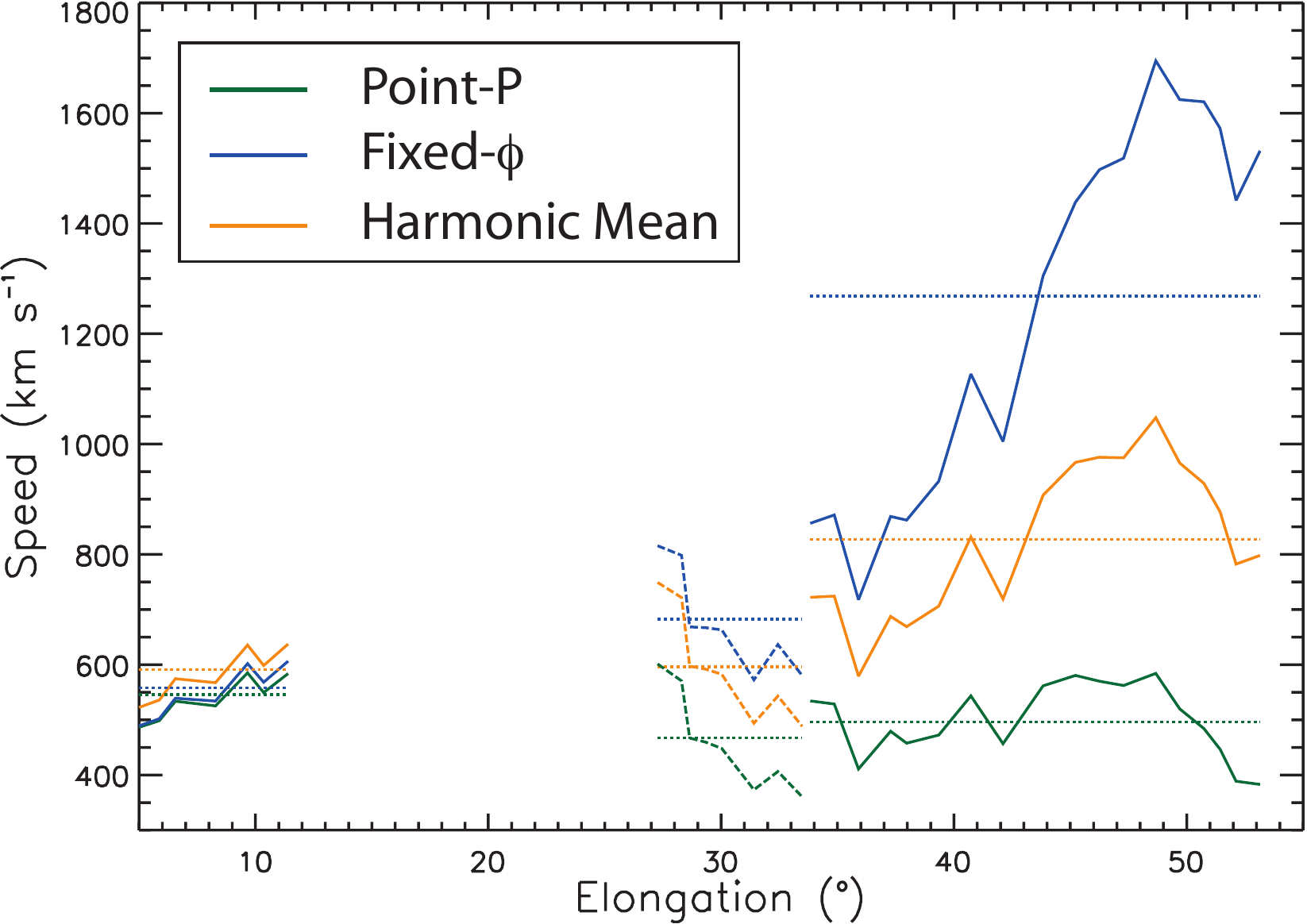}
\end{center}
\label{fig:position}
\caption{Position ({\it top}) and speed ({\it middle} and {\it bottom}) of the two fronts at PA 69 according to the three methods. The errors are typically 1.5$\%$ for the position and 15$\%$ for the speed. The averages are shown with dotted lines and the second front with dashed lines.}
\end{figure}

\subsection{Data analysis}
With this method, we re-analyze the data from the two fronts observed by HI-2 on  January 26-27, 2007. We analyze the data at PA 69, where the SECCHI's coverage is best for this event. %because it corresponds to the apparent central PA of the SECCHI's instruments (due to the roll of STEREO). 
There were only limited observations of the two ejections prior to the data gap. For the first ejection, all three methods agree and give an average speed between 550 and 600~km~s$^{-1}$ at 40~R$_{\odot}$, which is consistent with LASCO observations and also with the speed of 604~km~s$^{-1}$ reported by \citet{Harrison:2008} for the front at PA 90 (i.e. the nose of the ejection).  For the second eruption, we use LASCO data, which give a speed of approximatively 1,200-1,300~km~s$^{-1}$ at 20~R$_{\odot}$. Next, we analyze the two fronts after the data gap in HI-2 FOV. The top panel of Figure~3 shows the derived position for the two fronts according to the three methods. First, it is worth noting that the F$\phi$ and the HM approximations differ by less than 10$\%$ up to approximatively 180~R$_\odot$ (40$^\circ$), but the HM approximation does not result in a large apparent acceleration at very large elongation angles. %The predicted arrival time of the nose of the CME at 1~AU between the two methods differ by about 6~hours (which is about 15$\%$ of the total predicted transit time).
Next, we derive the speed of the two fronts according to these methods. We plot a running average over approximatively 5 hours to reduce the magnitude of the error in the speed. HI-2 resolution is 4~arcmin; assuming the elongation angles are measured with a precision of 5~pixels, the error in position is of the order of 1.5$\%$ and the resulting speed has an error of about 15$\%$.

We believe the analysis of the numerical results from section 3.2 shows that the PP method cannot be used to study the speed of limb CMEs past 100~R$_\odot$, which is the approximate position of the two fronts after the data gap. According to the F$\phi$ and HM methods, the second front, which fades out at about 33$^\circ$ elongation, has an average speed of 680 and 605~km~s$^{-1}$, respectively, with a general decelerating trend with an initial speed around 750-850~km~s$^{-1}$ around 100~R$_\odot$. The two methods are overall consistent with each other, and we believe this shows that the transient associated with the second front had a speed of 750-850~km~s$^{-1}$ around 100~R$_{\odot}$ and decelerated to 500-600~km~s$^{-1}$ before disappearing around 140~R$_{\odot}$.
 
\begin{table}[htp]
\caption{Summary of the speeds measured by SECCHI for the two fronts.}
\vskip4mm
\centering
\begin{tabular}{ccc}
\tophline
Front & Speed Before Collision & Speed After Collision\\
\middlehline
1 & 600~km~s$^{-1}$ & 850-900~km~s$^{-1}$\\
\middlehline
2 & 1200-1300~km~s$^{-1}$ & 800~km~s$^{-1}$ @ 80~R$_\odot$ \\
 & &  550~km~s$^{-1}$@ 140~R$_\odot$\\
\bottomhline
\end{tabular}
\end{table} 
 
For the first front, which is tracked until 53$^\circ$, the F$\phi$ results in a strong acceleration after 40$^\circ$ elongation and the PP method in an almost constant low speed. Fast CMEs are not expected to experience large acceleration in the heliosphere \citep[e.g.][]{Gopalswamy:2001,Tappin:2006}. The HM method results in a speed more consistent with this fact than the F$\phi$ method, although it shows a limited, unphysical acceleration at large elongation angles. The average speed obtained from the three methods is 490, 1340 and 845~km~s$^{-1}$, respectively; the average speed of the F$\phi$ and HM methods for observations between 28$^\circ$ and 40$^\circ$ is 880~km~s$^{-1}$ and 705~km~s$^{-1}$. The analysis is more complicated than for the second front, but, we believe that the observations are consistent with a transient whose average speed is about 850-900~km~s$^{-1}$ (the average value of the HM method, and the average value of the F$\phi$ within 40$^\circ$). 

\subsection{Consequence for the process of CME-CME interaction}

\begin{figure*}[t]
\vspace*{2mm}
\begin{center}
\includegraphics[width=12cm]{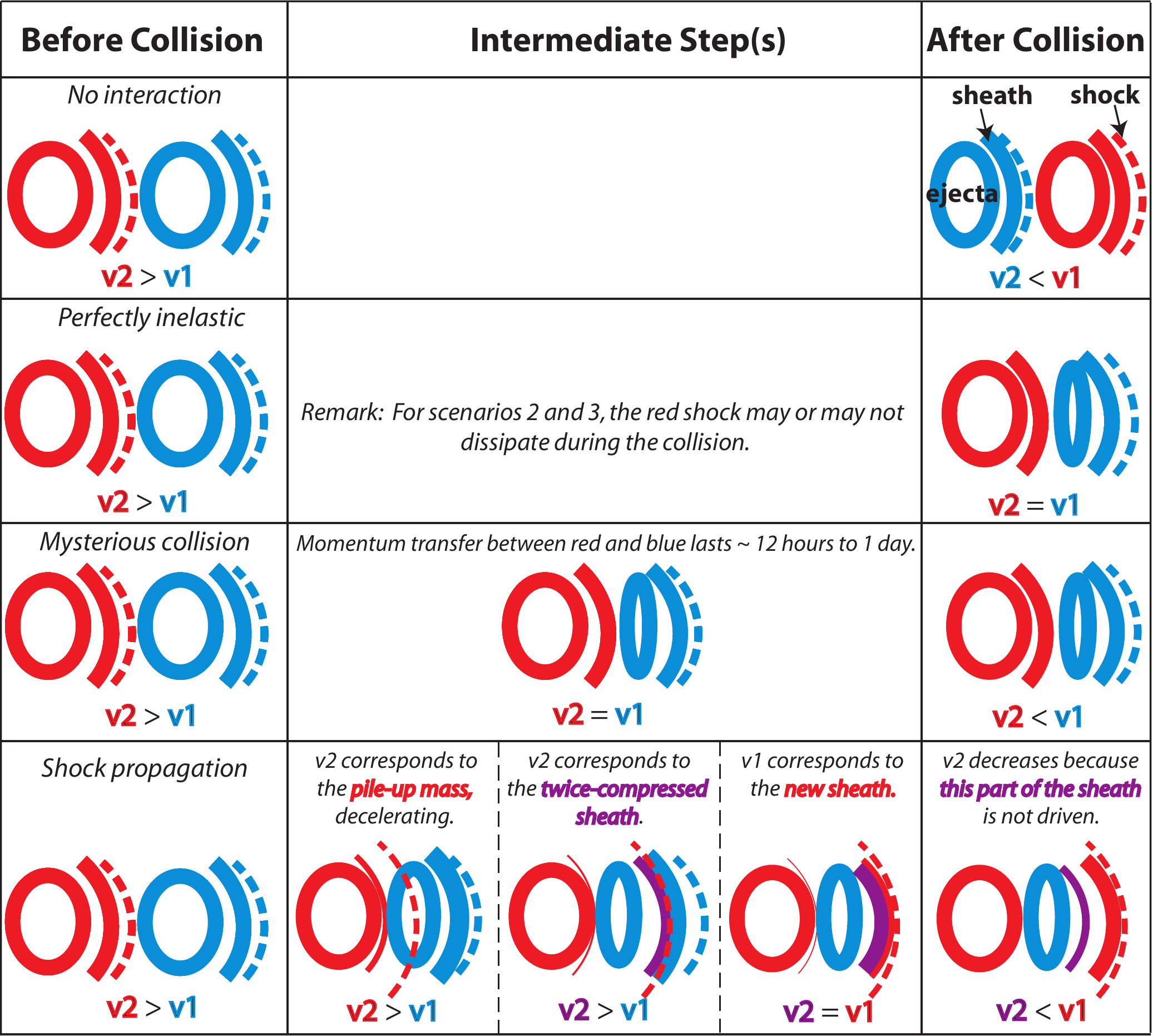}
\end{center}
\caption{The four scenarios for CME-CME collisions that might explain the fact that the first front after the collision is faster than the second front. In the sketches, the ellipses, the solid arcs, and the dashed arcs correspond to the ejecta, dense sheaths and to the shock waves, respectively.}
\label{fig:scena}
\end{figure*}

The derived speeds of the fronts are summarized in Table~1. We believe there are 4 scenarios consistent with the result that the first front is faster than the second front after the data gap;  we analyze these scenarios with respect to the measured speeds of the two fronts. A schematic view of the 4 possibilities is shown in Figure~4. %We argue that the SECCHI measurements are consistent with a momentum exchange associated with the second (faster) shock wave propagating through the first (slower) ejecta and sheath. 
In the first scenario, the January 25 CME could have ``passed'' the January 24 CME without major interaction. This scenario is possible if the two eruptions have a large angular separation, and if they do not propagate along the same direction. Then, part of the fast front could, in the projected images, ``pass'' the slow front when in fact there is no interaction. This scenario is described in greater details in \citet{Webb:2009}. While it is plausible that only a small part of the two CMEs interacted and that the major part of the January 25 CME simply passed next to the January 24 CME without interaction, we believe this is very unlikely. First, it is hard to understand how the speed of the January 24 eruption could be faster after the data gap than before; also, the January 25 eruption shows a strong deceleration during the data gap, which tends to suggest some form of interaction. Second, the measured width of the eruptions --greater than 100$^\circ$ in LASCO FOV as reported in \citet{Webb:2009}-- also makes a missed encounter implausible. Last, this is not supported by any MHD models, which tend to show that CMEs act as magnetic barriers. This scenario could however explain what happened if the two CMEs were associated with different active regions and, consequently, had a large angular separation. This separation could be as large as 35$^\circ$ if the first CME was associated with the eastern most active region and the second CME with the western most active region present in January 24-25, 2007. Our arguments to associate both ejections with the same active region can be found in \citet{Lugaz:2009b}. 

In the second and third scenario, the two CMEs collide, the collision is associated with momentum transfer between the ejections (as \citet{Farrugia:2004} considered). The observations appear to be consistent with both eruptions having the same speed after the collision, i.e. a perfectly inelastic collision. However, it is hard to understand the evolution of the speed of the two CMEs after the collision according to this scenario. If the January 25 CME pushes the January 24 CME, both fronts should have a similar speed at all times after the collision. This scenario appears more plausible if one believes the speeds derived using the PP method. However, using the PP speeds and positions, the average transit speed of the two fronts during the data gap should be 500 and 650~km~s$^{-1}$ respectively. This scenario would therefore be consistent with a large deceleration of the January 25 (fast) CME and almost no acceleration of the January 24 (slow) CME, which, in turn, can only happen if the January 24 CME is much more massive than the January 25 CME. \citet{Webb:2009} reported the mass of the January 24 and 25 CMEs being  $4.3 \times 10^{15}$~g and $1.6 \times 10^{16}$~g, respectively, making this scenario very unlikely. 

In the third scenario, the collision is elastic and there is a momentum transfer from the second to the first ejection on a time-scale of 12-20~hours. The momentum transfer has an unknown cause and continues until the second eruption becomes slower than the first one. This scenario is not fundamentally different from the last one, which does not require unknown processes and can explain the disappearance of the second front.

In the fourth scenario, the unknown process is, in fact, the compression and momentum transer associated with the shock wave from the January 25 CME. Before the CMEs collide, the shock wave driven by the January 25 CME propagates through the January 24 CME (ejecta and sheath), compressing and accelerating it, before merging with its associated shock wave. After the data gap, the first front corresponds to the sheath associated with the merged shocks. Due to its interaction with the January 24 CME and sheath, the shock wave initially associated with the January 25 CME has decelerated rapidly to a speed $\sim$~850~km~s$^{-1}$. There are two possibilities to explain the second front: it could be the remnant of the sheath associated with the January 25 shock wave which is ``trapped" between the two CMEs and ``forced'' to propagate with a speed comparable to that of the January 24 CME. However, the distance between the two fronts is of the order of 20~$R_\odot$ at PA 69. 
If the January 24 CME is between the two fronts, this would mean that the magnetic cloud has been compressed to less than 20~$R_\odot$, which does not seem reasonable. Moreover, the two fronts appear to merge along PA 90, which is inconsistent with this explanation. The other possibility is that it is associated with a transient phenomenon during the shock-CME or shock-sheath interaction or three-dimensional effects. For example, the core (or any part of the cloud) of the January 24 CME could be have been compressed and accelerated by the shock wave and it relaxes to slower speeds. 
Most likely, part of the sheath associated with the January 24 CME gets compressed to very high density and relaxes to the average value of the new sheath \citep[similar to what has been discussed in][]{Lugaz:2005b}. %Or, this might be a geometrical effect, where the interaction between the January 25 shock wave and the January 24 sheath (which is expected to result in very large densities) is observed to happen at different times along different longitudes. 
However, each of these sub-scenarios involve the propagation of the January 25 shock through the January 24 CME. 
We note that this scenario does not require the presence of a shock wave driven by the January 24 CME, but simply a sheath of dense material (piled-up mass and/or compressed material) ahead of the CME. The only difference due to the possible absence of the first shock wave is that there is no instance of shock-shock merging. Therefore, the shock wave ahead of the merged CMEs after the interaction is simply the shock wave originally driven by the January 25 CME now propagating into an unperturbed solar wind. 

\conclusions[Discussions and Conclusions]

In the first part of this study, we have tested the two most common methods used to derive CME radial distances from elongation angle measurements, the Point-P and Fixed-$\phi$ methods. Confirming previous work by \citet{Kahler:2007} , \citet{Wood:2009} and \citet{Webb:2009} we find that, above 35$^\circ$, both methods yield poor results, especially for CME speed and acceleration. We propose an alternative analytical method to derive CME radial distances. %To reduce the number of free parameters, 
We consider a very simple model, namely that the density peak maps out as a sphere whose center propagates radially outward from the flare location, and that the elongation angle corresponds to the angle of the line-of-sight tangent to this sphere. We find that the diameter of this sphere is given by the {\bf harmonic average of the Point-P and Fixed-$\phi$} approximations further corrected by 1/$\cos(\mathrm{PA_{app}})$ where $\mathrm{PA_{app}}$ is the position angle with respect to the nose of the CME. For a limb ejection, this method gives results similar to the Fixed-$\phi$ approximation up to about 40$^\circ$ and more realistic results at larger elongation angles. The Point-P and Fixed-$\phi$ approximations are expected to yield a lower and upper-bound to the actual distance of a CME \citep[e.g., see][]{Webb:2009}. Any alternative methods to determine radial distances from elongation angles shall fall in-between, as is the case here. However, we find a particular physical interpretation for the harmonic mean of these two methods. We have also found a new interpretation of the position derived from the Fixed-$\phi$ approximation, namely that it is the diameter of the sphere representing the CME if the emission is assumed to originate directly from the Thomson sphere. This might explain why this approximation works fairly well even for wide CMEs.

We must be aware of the limitations of this method. First, this is most appropriate for wide CMEs such as the ones observed in January 24-25, 2007, because the assumed geometry is consistent with a CME whose angular extent is 90$^\circ$. This approximation, while arbitrary, is required to reduce the number of free parameters of the model to one. 
It also appears to be a better approximation for wide CMEs than the Point-P approximation which is consistent with a CME whose angular extent is 360$^\circ$. Secondly, this model assumes that the CME propagates on a fixed radial trajectory, ignoring heliospheric deflection. This is the same assumption made to derive the Fixed-$\phi$ approximation and is also required to reduce the number of free parameters. In future works, we shall investigate how stereoscopic observations of CMEs by the two STEREO spacecraft can help relax these two conditions. 
Last, the model of CMEs use to derive this approximation assumes that the CME front (part piled-up mass, part shocked material) maps out as a sphere. As noted in the introduction, CME fronts are known to be distorted and usually flattened from their interaction with the structured solar wind. In Figure~1, we have shown two simulated instances where this approximation is more or less appropriate; it is worth noting that the two simulations use different models of CME initiation and solar wind. Assuming a more complex shape (for example a ``pancake'') would require a fitting of the model and could not yield a direct analytical relationship such as the one derived here.

We have re-analyzed the HI-2 measurements in January 26-27, 2007 associated with two interacting CMEs with the three methods. We found that the first bright front after the interaction corresponds to a transient propagating with a speed of about 850~km~s$^{-1}$, while the second front corresponds to a transient whose speed decreases from 850~km~s$^{-1}$ to 550~km~s$^{-1}$ in about 12-18~hours before ultimately disappearing. Among the 4 scenarios which could explain that the acceleration of the first front relative to the second one, we found that the only scenario consistent with the observations require that the part of the shock wave driven by the (faster) January 25 CME propagates first through the (slower) January 24 CME. The propagation of this fast shock wave inside the CME and dense material of the sheath results in its large deceleration. The most likely explanation for the origin of the second front is that it is part of the dense sheath associated with the January 24 ejection, which is compressed and accelerated by the shock and decelerates to the speed of the January 24 magnetic cloud. 

Our analysis has been limited to one case of a fast, wide limb CME. It is for this particular geometry that the Fixed-$\phi$ approximation is expected to give the largest error at large elongation angles. However, we believe our new average method should provide an improvement over the existing methods, notably over the Point-P approximation for wide eruptions, and that it should be used complimentary to three-dimensional fitting methods and numerical simulations. We plan to test and validate this relation for other heliospheric observations of wide and fast CMEs, starting in the near future with the April 26, 2008 eruption. Our analysis of the January 24-27, 2007 observations is the first heliospheric observational evidence of a shock wave propagating inside a CME. More observations without data gaps are required before we have a more definite understanding of CME-CME interaction.

\begin{acknowledgements}

The research for this manuscript was supported by NSF grants ATM-0639335 and ATM-0819653 as well as NASA grant NNX-08AQ16G. 
  We would like to thank the reviewers, Tim Howard and an anonymous referee for helping us to improve and to clarify this manuscript.
   The SECCHI data are produced by an international consortium of 
  Naval Research Laboratory, Lockheed
  Martin Solar and Astrophysics Lab, and NASA Goddard Space Flight
  Center (USA), Rutherford Appleton Laboratory, and University of
  Birmingham (UK), Max-Planck-Institut f{\"u}r Sonnensystemforschung
  (Germany), Centre Spatiale de Liege (Belgium), Institut d'Optique
  Th{\'e}orique et Appliqu{\'e}e, and Institut d'Astrophysique
  Spatiale (France).  SOHO is a project of international cooperation between ESA and NASA,  and the SOHO LASCO/EIT catalogs are  maintained by NASA, the Catholic University of America, and the US Naval Research Laboratory (NRL).

\end{acknowledgements}

\end{document}